\documentclass[manuscript]{emulateapj}

\slugcomment{Submitted to The Astrophysical Journal Letters}

\shorttitle{{\em Kepler} Short Cadence Data Characteristics}
\shortauthors{Gilliland, et al.}

\begin{document}


\title{INITIAL CHARACTERISTICS OF {\em KEPLER} SHORT CADENCE DATA}


\author{Ronald L. Gilliland\altaffilmark{1},
Jon M. Jenkins\altaffilmark{2}, William J. Borucki\altaffilmark{3},
Steven T. Bryson\altaffilmark{3}, Douglas A. Caldwell\altaffilmark{2},
Bruce D. Clarke\altaffilmark{2}, Jessie L. Dotson\altaffilmark{3},
Michael R. Haas\altaffilmark{3}, Jennifer Hall\altaffilmark{4},
Todd Klaus\altaffilmark{4}, David Koch\altaffilmark{3}, Sean McCauliff\altaffilmark{4},
Elisa V. Quintana\altaffilmark{2}, Joseph D. Twicken\altaffilmark{2},
and Jeffrey E. van Cleve\altaffilmark{2}}
\altaffiltext{1}{Space Telescope Science Institute, 3700 San Martin Drive,
Baltimore, MD 21218}
\email{gillil@stsci.edu}
\altaffiltext{2}{SETI Institute/NASA Ames Research Center, MS 244-30, Moffett Field, CA 94035}
\altaffiltext{3}{NASA Ames Research Center, MS 244-30, Moffett Field, CA 94035}
\altaffiltext{4}{Orbital Sciences Corporation/NASA Ames Research Center, MS 244-30, Moffett Field, CA 94035}



\begin{abstract}
The {\em Kepler Mission} offers two options for observations -- either
Long Cadence (LC) used for the bulk of core mission science, or Short Cadence
(SC) which is used for applications such as asteroseismology of solar-like
stars and transit timing measurements of exoplanets where the 1-minute
sampling is critical.  We discuss the characteristics of SC data obtained
in the 33.5-day long Quarter 1 (Q1) observations with {\em Kepler} which completed
on 15 June 2009.  The truly excellent time series precisions are nearly Poisson limited at
11th magnitude providing per-point measurement errors of 200 parts-per-million
per minute.  For extremely saturated stars near 7th magnitude precisions of
40 ppm are reached, while for background limited measurements at 17th magnitude 
precisions of 7 mmag are maintained.  We note the presence of two 
additive artifacts, one that generates regularly spaced peaks in frequency,
and one that involves additive offsets in the time domain inversely 
proportional to stellar brightness.
The difference between LC and SC sampling is illustrated for transit observations
of TrES-2.
\end{abstract}




\section{INTRODUCTION}

The {\em Kepler Mission} has a primary science
goal of detecting analogs of the Earth orbiting stars in the extended
solar neighborhood as reviewed in \citet{bor09}.
Overall mission design and performance are reviewed by \citet{koc09}.
In this paper we establish the characteristics
of data from {\em Kepler} collected at 1-minute cadence.  Asteroseismology
of solar-type stars requires detection of oscillations with periods of 3 -- 10
minutes, which thus cannot be studied at the LC of nearly 30 minutes.
Early science returns and goals of asteroseismology are reviewed by \citet{gil10}.
For exoplanets,
precise determination of the transit times of many successive events to search
for possible variations in those times due to the gravitational
influence of other planets (\citealp{hol05}) benefits from SC.

\section{TARGET, DATA, AND CADENCE SPECIFICS}

The {\em Kepler} targets in Q1 consisted of 156097 stars observed
at LC.
A mere 512 targets, or 0.3\% of the total may be 
carried at the roughly one minute SC.  A new SC target list may be 
used each month.
Early in the mission nearly all the SC targets
have been assigned to the {\em Kepler Asteroseismic Consortium} (KASC);
some 4000 targets will be surveyed, each for one month from which the
best set will be selected for extended observations.  As planetary
candidates are discovered many of these are switched to short 
cadence in order to provide precise transit timing variation (TTV)
measures, and for accurate determinations of the host star mass and 
radius if stellar oscillations can be detected.
The asteroseismology program through KASC is guaranteed access
to 140 SC target slots throughout the mission. 
At the start of Q2 (19 June 2009) 25 SC targets are reserved for the Guest Observer (GO) program.

There is
also a restriction that the total number of
pixels for SC not exceed 512$\times$85 = 43,520. 
Very bright stars require more than 85 pixels each, thus 
requiring an SC target set with a balanced distribution of magnitudes.

Data is time-stamped so that the mid-time of each cadence is
known with an accuracy of $\pm$ 0.050 seconds.
Details of how the data acquisition process is
managed on-board the spacecraft may be found in the ``{\em Kepler} Instrument
Handbook" \citep{van09}, as well as a wealth of general 
information that is beyond the scope of a {\em Letter}.
\citet{cal09} discuss characteristics of the instrument and early calibration.

At the most basic level it is important to understand that the integrations
underlying both SC and LC data are the same 6.02 s, 
followed by 0.52 s readout, before successive integrations are summed
into memory.
There is no difference at which saturation
occurs for the 58.84876 s cadence data, versus the 29.4244 min LC, since both
are based on 6 s exposures.
Within the 58.85 s SC periods, 54.18 s is spent usefully collecting photons
while the remainder of the time is divided between 9 readouts, which 
generates ``smear" along the columns as {\em Kepler} has no shutter.
This is removed \citep{jen09a} as an early pipeline calibration step.
In Q1 138 out of 49,170
cadences ($<$0.3\%) were rejected, being associated with periods
of excess spacecraft jitter yielding a 99.7\% duty cycle (!) for these 33.49 days, which span HJD (-2454900) =
64.00 to 97.49.

LC data involve collection of collateral data
continuously in a spatial sense along overscan (trailing black) columns,
plus masked and virtual (smear) rows,
while for SC only those collateral pixels that project onto the SC
aperture are retained.  Likewise, for LC there are 4464 pixels on each
channel assigned to apparently blank regions of the field to provide real
time monitoring of sky brightness (which changes more than expected),
SC does not have dedicated sky pixels, therefore
interpolation between LC measures is used by the current pipeline processing.
Sky was not expected to change on a time scale of minutes, and
inclusion of direct background monitoring at SC would have 
increased telemetry needs for SC dramatically leading to an
early decision to forgo this.

\section{ERROR BUDGET, AND REALIZED NOISE LEVELS}

In pre-launch simulations dozens of factors were considered in combination
to arrive at predictions of noise levels in the {\em Kepler} data.
In this {\em Letter} we consider the
minimum number of error budget terms that together provide a good
empirical representation of the realized noise level.  
With {\em Kepler's} exquisite precision most
stars are variable at least on time scales of days.
Detrending of photometric time series against variations of $x, y$
pointing, thermal and focus changes is discussed by
\citet{jen09b} for LC data.  For SC applications where the time scales
are usually short compared to thermal and focus forcing functions this is less of
an issue, but still one that will remain a major analysis factor going
forward.

The pipeline \citep{jen09a}
provides ``raw" time series which are the simple sums
(after cosmic ray and background removal) over the pixels
chosen in an aperture to provide optimal signal-to-noise (S/N), the
units of these are detected electrons (e-) per 58.8 s SC.
To turn these time series into relative photometry with a mean of zero,
and remove slow trends
the following steps for each star have been taken:  (1) A 5th order 
polynomial is fit to the full 33.5 days of data
to remove drifts on timescales of a few days to a month resulting
from minor pointing drift, focus changes etc.  Future releases 
from the pipeline are expected to have more effective detrending,
perhaps obviating any need for this step.  (2) The polynomial
fit is subtracted and then this value divided by the zero point term of the polynomial
yielding zero mean, empirically detrended data.  (3) A running median filter
361 cadences wide (6 hours) is evaluated and point-by-point subtracted 
from the data.
SC data processed
before November 2009 had an underlying software error that resulted in blocks of 3,000
successive pixels having photometric values too high by $\sim$0.3 mmag
at 12th magnitude (scaling inversely with brightness).  These
randomly positioned blocks averaged about one per star, are 
effectively suppressed in the current data by the median filter,
and the software error has now been fixed in the pipeline.
(4) The \S 4.1 correction is applied.
(5) Any individual
points more than 5-$\sigma$ deviant from the mean of zero have
values and weights set to zero.  Typically this final
step clips values for $<$0.05\% of the data points.

Fig. 1 shows the result of forming power spectra for all 512 SC targets
up to 8 mHz with a spacing of 0.05 $\mu$Hz that over-samples the 
frequency resolution by a factor of about seven, after transforming this to
amplitude with an appropriate scale factor to have units of ppm.
We then measure the noise level (standard deviation of the amplitude 
spectrum) over $\nu$ = 1 -- 7.5 mHz.  Many stars have strong variations below
1 mHz, relatively fewer show significant power at higher frequencies,
although a healthy subset of intrinsically variable stars show excess power at high $\nu$.
(10 stars
show high noise due to poorly designed apertures
early in the mission.)  Evident in Fig. 1 is a lower envelope
of data points representing the
empirical limit to precision as a function of magnitude.  

\begin{figure}[h!]
\resizebox{\hsize}{!}{\includegraphics{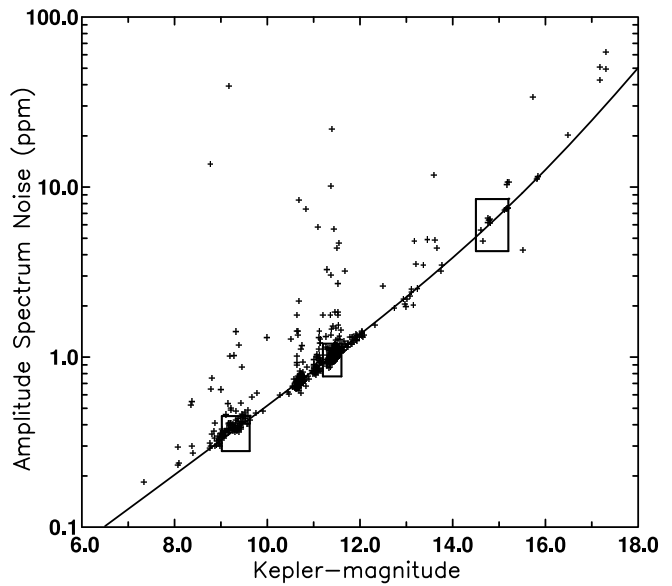}}
\caption{The noise level for each of 512 stars in Q1 Short Cadence
data expressed as the standard deviation in ppm over 1.0 -- 7.5 mHz
in an amplitude spectrum (square root of power).  The curve passing
through the lower envelope consists of simple
Poisson, readout noise plus sky background terms.  The boxes
near magnitudes of 9.3, 11.4 and 14.9 respectively are used to
delineate samples of quiet stars as discussed in the text and Fig. 2.
A minor spread in values is expected from differences
in sensitivity and readout noise over the 84 amplifiers as well as
errors in the {\em Kepler Input Catalog} magnitudes.}
\label{fig:01}
\end{figure}

The Fig. 1 lower envelope is fit ``by-eye" with
$10^6 ({\tt c} + 9.5 \times 10^5 (14.0/Kp)^5)^{1/2}/({\tt c} N^{1/2})$
where {\tt c} = 1.28 $\times 10^{(0.4(12. - Kp) + 7)}$ is the number of detected e- per cadence
and $N$ = 49032 (the number of SCs used in Q1).
In this empirical formulation aimed at achieving a minimal term 
model for the noise, {\tt c} provides the Poisson term and is within
the small channel-to-channel calibrated value for the count rate,
$Kp$ is the {\em Kepler} magnitude (close to $R$-band in center wavelength,
but much broader, e.g. see \citealp{koc09}), and the
$(14.0/Kp)^5$ term captures to first order the number
of pixels used in the aperture extraction as a function of magnitude.
We have assumed 8-pixel apertures at $Kp$ = 14, increasing to 256
at $Kp$ = 7.  This in turn implies a per pixel variance of 11,875
per cadence, or a readout plus Poisson statistics on the sky
of 109 e- per pixel per underlying 6 second exposure (9 per cadence).
Since this is approximately the expected value \citep{cal09} we conclude that this
minimal model successfully represents the noise (as measured at high frequencies from amplitude spectra)
in the data.

The set of 115 quiet stars at $Kp$ = 11.44
has a mean time series {\em rms} of
255 ppm, and a mean amplitude noise level of 1.03 ppm.  (The mean
{\em rms} value divided by $N^{1/2}$ is 1.15 ppm, which is larger than
the quoted amplitude spectrum noise level, since only the {\em rms} includes
contributions from below 1 mHz.)

As a simple rule of thumb reaching a noise level of 1 ppm is 
required to make the most basic asteroseismic 
measurement, that of determining the so-called large splitting
$\Delta \nu_0$ which in turn can be used to accurately constrain the
mean stellar density (see, e.g. \citealp{gil10})
in solar-type stars where several individual
modes may be expected to have amplitudes of a few ppm.
We reach this level with one month of data at $Kp$ $\sim$ 11.3.
Coincidentally, $Kp$ $\sim$ 11.3 is also the rough dividing line below
which stars saturate the detector during the 6 s integrations.
It had long been a goal with {\em Kepler} 
to achieve photometry
near the Poisson limit on strongly saturated stars as had been demonstrated
for multiple {\em HST} CCD-based instruments
without this having been a design goal
(see, e.g. \citet{gil04} for ACS).
Nearly Poisson limited photometry is maintained to 
$Kp$ = 7 (factor of 50 over-saturated), and beyond as long
as the bleed does not extend into columns beyond the edge of the detector,
nor blend with other bright stars.  Photometry of highly
saturated stars merely requires using an aperture that consistently 
captures the super-set of pixels that are bled into at any time over the
time series.
{\em The Kepler short cadence data have demonstrated a large dynamic
range of 10 magnitudes, and for special applications can probably be extended
usefully by one magnitude in each direction beyond the 7 -- 17 range.}

A major factor contributing to the stability of {\em Kepler}
photometry is the operating environment of the Earth-trailing orbit.
The 2-axis, point-to-point jitter in SC centroids after removing
a 5th order polynomial in $x$ and $y$ separately is 4.2 $\times 10^{-4}$
pixels, or only 1.7 milli-arcseconds.  This remarkable result
holds despite the use of two variable guide stars (removed for Q2)
known to have degraded guiding in Q1 \citep{jen09b}.

\section{KNOWN ANOMALIES IN CURRENT SC TIME SERIES}

A number of anomalies, any specific one of which is a surprise, appear
in the early {\em Kepler} SC data.  The fact that some such unanticipated
anomalies have arisen should not of course come as a surprise.
We discuss the two most significant examples of these.

\subsection{Power at 1/LC-period and Overtones}

Fig. 2 illustrates the most serious artifact appearing in power
spectra of SC time series.  A number of peaks corresponding to the fundamental
and all harmonics of the inverse of the LC-period, 29.4244 minutes,
contaminate the spectra.  Comparison of the amplitude of
these modes shows that the effect is additive -- as can be seen
in Fig. 2 the relative amplitudes scale inversely with the 
stellar intensity.  The causal factor may be variation in the 
detector electronics that ``rings" in response to changes associated
with initiation of each LC.  For a given star these 
sinusoids appear to be coherent over the full 33 days, but different
phases apply to each star.  These peaks also appear in power spectra of the 
overscan (black) pixels.  Attempts to ameliorate this effect, e.g.
by using local sky variations from the periphery of 
SC target apertures, have failed to improve the situation.
Further study will determine if changes to pipeline 
calibration can suppress this contamination which is introduced
on the spacecraft.  The SC photometry folded
on the LC period (not shown) has significant ringing that
is low amplitude near the boundaries, and larger peaks near
the center at a time scale of 3 -- 4 short cadences thus explaining
the larger strength of intermediate harmonics,

\begin{figure}[h!]
\resizebox{\hsize}{!}{\includegraphics{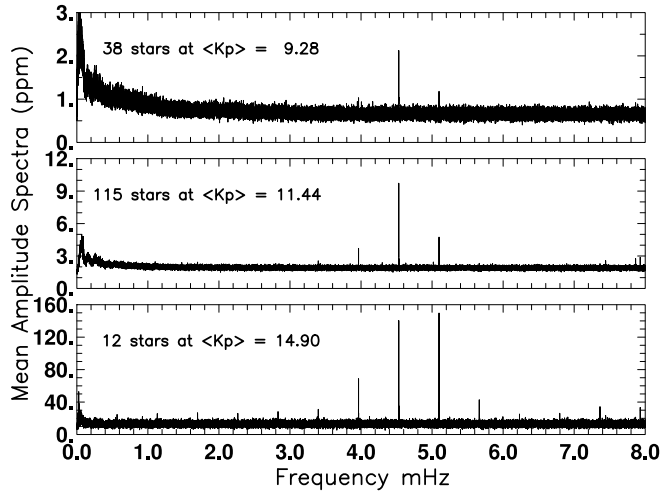}}
\caption{Mean amplitude spectra over samples of quiet stars
spanning more than a factor of 100 in brightness are shown.
The 1/LC-cadence artifacts at the fundamental of 0.566391 mHz
and all harmonics are visible for the faint star set in the
bottom panel.  Even at 9th magnitude in the upper panel this
artifact remains a dominant spectral feature from the 7th and
8th harmonics.}
\label{fig:02}
\end{figure}
whether this robust empirical behavior is significant in 
understanding the origin remains under investigation.

For now, at least, asteroseismic investigations will either need
to attempt corrections in the time or frequency domains on a star-by-star
basis, or at a minimum flag frequencies at multiples of 0.566391 mHz
as suspect.  Since the fraction of ``phase space" contaminated by 
these peaks is roughly the number of harmonics (14) times the 
frequency resolution (1/33-days), divided by 8 mHz, which equals
0.06\%, the ultimate impact of these spurious peaks should be minimal.

With the exception of this artifact the {\em Kepler} spectra are
remarkably clean and free of problematic sidelobes.
The low frequency turn up in power for 9th magnitude stars below $\sim$2 mHz
in Fig. 2 likely follows from stellar oscillations and granulation noise 
which are easily seen in many individual amplitude spectra for bright stars.
The peak at $\sim$0.1 mHz likely corresponds to an artifact
near 3.2 hours discussed in \citet{jen09b};
the wavy structure at low frequency in the middle panel is
likely due to harmonics of this.

\subsection{Additive Corrections in the Time Domain}

A somewhat more problematic artifact is illustrated in Fig. 3, but
unlike that from the previous section, 
this can likely be addressed with pipeline modifications.

\begin{figure}[h!]
\resizebox{\hsize}{!}{\includegraphics[scale=1.0]{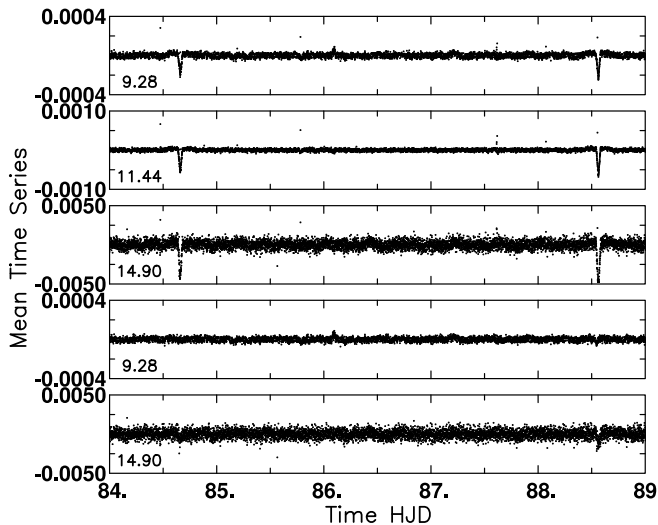}}
\caption{The upper three panels show means, after sigma clipping, over the same
sets of stars shown in Fig. 2; this time however for 5 days
of the direct time series.  Events in the pipeline provided
time series at HJD = 84.7 and 88.6 scale inversely with 
stellar brightness.  The lower two panels show corrected
mean time series for the 9th and 15th magnitude bins as
discussed in the text.}
\label{fig:03}
\end{figure}

The primary offsets visible in Fig. 3 are vaguely transit-like, and
clearly scale inversely with the intrinsic stellar
intensity.
These appear at the times of the two major brightening events seen
in Q1 LC sky background (see \S 3 of \citealp{jen09a} for a discussion
of these enigmatic and unexpected events) for which imperfect background
correction is applied in the pipeline for these SC data.
The correction illustrated adopts the mean from 115
stars near $Kp$ = 11.44, and for points deviating by more than $8 \times 10^{-5}$
from the mean of zero subtracts a scaled version of this.
While not a 
perfect resolution, the primary features largely disappear, as
well as several positive deviations appearing in the time series.

\section{SCIENCE APPLICATIONS OF SC TIME SERIES}

\subsection{Detailed Study of Transits}

The previously known exoplanet hosts, TrES-2 \citep{odo06}, HAT-P-7 \citep{pal08}, and
HAT-P-11 \citep{bak09} have all been observed at SC from the start.
TrES-2b has the narrowest transits of these and is used to 
illustrate differences between LC and SC data in Fig. 4.

\begin{figure}[h!]
\resizebox{\hsize}{!}{\includegraphics[scale=1.0]{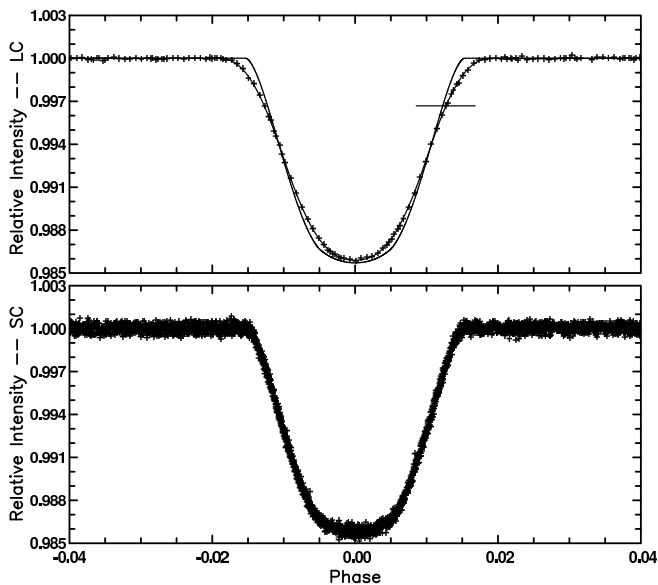}}
\caption{The upper panel shows the 33.5 days of Q1 LC data
folded on the $\sim$2.47 day period of TrES-2 with a {\em Kepler} magnitude
of 11.34.  The lower panel shows folded data at SC
for the same interval.  The curve passing through the points
in the upper panel was based on fitting a transit light curve
model {\em integrated} over the LC of 29.4 minutes, while
for comparison the narrower curve (at top) shows the fit to SC
data.  The horizontal bar in the upper panel shows the length
of an LC interval in phase for TrES-2.}
\label{fig:04}
\end{figure}

LC observations effectively integrate over 29.4 minute intervals,
which are not short compared to features in the TrES-2b transit,
and therefore exhibit easily discerned differences with respect to
SC data which much more finely sample the light curve.
The two fits shown in Fig. 4 use the \citet{man02} transit
model with identical values assumed for stellar mass, radius and
limb-darkening.  Parameters solved for are:
$(R_p/R_*)^2$, the orbital period and phase, a zero point outside
of transit, and a transit width related to the impact parameter.
The SC model fit ({\em rms} residuals are 237 ppm)
evaluates the functional representation
at each central time, while the LC model 
fit ({\em rms} of 66 ppm)
is integrated over the LC intervals
(using a numerical sum corresponding to the 30 contributing SC intervals).
Although the blurring introduced by the long LC intervals
is very noticeable for TrES-2, the formal parameters returned
from a simple non-linear, least-squares fit are nearly identical
in the two cases.  The solutions for planetary radius differ
by only 0.012\% for the LC and SC solutions, while the inferred
orbital inclinations differ by only 0.001 degrees.  Since systematic
uncertainties on the stellar radius will likely remain $>>$ 0.01\%,
even if asteroseismology solutions for the mean stellar density
are available, LC data seem to equally well support determination
of the basic parameters associated with transit modeling.

SC data are expected to provide a significant relative advantage for transit
timings where LC data fail to resolve the critical ingress and
egress periods.  Quantification of this awaits applications to
longer time series generally needed for TTV analyses.

\subsection{Asteroseismology}

The case for short cadence data for asteroseismic applications is clear.
Indeed, the value of slightly less than 1-minute, rather than, say 2-minutes
for SC data which would have been more convenient in terms of on-board
data storage and telemetry bandwidth was selected to support robust
asteroseismology of solar-like stars some of which will have modes with periods as 
short as 2 minutes.
The Nyquist frequency for LC sampling of 283.2 $\mu$Hz permits 
asteroseismology on relatively large stars, e.g. the $R/R_{\odot}$ $\sim$ 6
star KOI-145 analyzed by \citet{gil10} has frequencies of 
maximum mode power (which scales with the surface acoustic cutoff frequency
$\propto \, {\rm g}/{\rm T}^{1/2}_{eff}$ -- \citealp{bro91})
at 143 $\mu$Hz and can be studied as effectively at LC as SC.
Thus near solar temperatures, stars with radii down to $R/R_{\odot}$ about 4
can be studied well at LC, for stars with radii below this SC data are needed.

\section{DISCUSSION AND SUMMARY}

We have shown that the 58.8 s short cadence data reaches {\em unparalleled} noise
levels near fundamental limits imposed by Poisson statistics on
source and background plus readout noise.
An annoying artifact imposes frequencies at harmonics of the LC 
period in power spectra, but these are restricted to a few 
known frequencies and will not generally degrade science applications.
SC data provide excellent returns in both transit and asteroseismic
analyses.  The SC data acquisition option is a limited resource
that will be in great demand throughout the {\em Kepler} mission.





\acknowledgments

Funding for this Discovery Mission is provided by NASA's Science Mission
Directorate.  We gratefully acknowledge the many individuals through whose
dedication and excellence the remarkable capabilities of {\em Kepler}
have been made possible.



{\it Facilities:} \facility{The Kepler Mission}.

\clearpage




\end{document}